\documentclass[epj]{svjour}
\usepackage{graphics}
\begin{document}

\title{Driving the resonant quantum kicked rotor via extended initial conditions}
\author{Alejandro Romanelli \and Guzm\'an Hern\'andez 
}                     
%
%
\institute{Instituto de F\'{\i}sica, Facultad de Ingenier\'{\i}a\\
Universidad de la Rep\'ublica\\
C.C. 30, C.P. 11000, Montevideo, Uruguay}
\date{Received: date / Revised version: date}
%
\abstract{ We study the resonances of the quantum kicked rotor
subjected to an extended initial distribution. For the primary
resonances we obtain the dispersion relation for the map of this
system. We find an analytical dependence of the statistical moments
on the shape of the initial distribution. For the secondary
resonances we obtain numerically a similar dependence. This allows
us to devise an extended initial condition which produces an average
angular momentum pointing in a preset direction which increases with
time with a preset ratio.\PACS{
      {37.10.Vz}{Mechanical effects of light on atoms, molecules
and ions}   \and
      {05.45.Mt}{Quantum Chaos semiclassical methods.}
     } 
} 
\maketitle
\section{Introduction}

\label{intro}

The quantum kicked rotor (QKR), can be considered a cornerstone in the study
of chaos at the quantum level \cite{CCI79}. Both in theoretical and
experimental terms, this topic is a matter of permanent attention \cite%
{Schomerus1,Schomerus2,Currivan,Sandgrove,Saunders,Halkyard}. An
experimental realization of the quantum kicked top (a variant of the QKR)
has been recently reported \cite{Chaudhury} and used to study the signatures
of quantum chaos. The behavior of the QKR has two characteristic modalities:
dynamical localization and ballistic spreading of the variance in resonance
\cite{Izrailev}. These behaviors are quite different and have no classical
analog. They depend on the relationship between the characteristic time of
the free rotor and the period associated to the kick. When the dimensionless
period of the kick, $T$, is an irrational multiple of $2\pi $ the average
energy of the system grows in a diffusive manner for a short time and
afterwards dynamical localization appears. In this case, the angular
momentum distribution is characterized by an exponential decay. When $T$ is
a rational multiple of $2\pi $ the behavior of the system is resonant with a
ballistic spreading. In this case the angular momentum distribution evolves
from the initial one in such away that its standard deviation has the time
dependence $\sigma (t)\sim t$. This quantum behavior is similar to the
optical Talbot effect \cite{Goodman,Patorski} in the time dimension.

The QKR has been experimentally realized through a dilute sample of
ultra-cold atoms exposed to a one-dimensional spatially periodic optical
potential that is pulsed on periodically in time (to approximate a series of
delta function kicks). The quantum resonances and the dynamical localization
have been experimentally observed in refs. \cite{Moore,Bharucha,Raizen,Kanem}%
. However with the aim of the present paper in mind, it is interesting to
highlight the work developed in refs. \cite%
{Currivan,Sandgrove,Saunders,Halkyard}, where the influence of an
initial momentum on the appearance of quantum resonances in the QKR
is explored experimentally establishing the dependence of the
quantum resonances on the initial velocity of the atoms. In
ref.\cite{Currivan}, the authors showed a sinusoidal dependence of
the energy on the initial momentum for two kicks, and a more complex
behavior with the same period at four kicks. They also proposed to
use the momentum dependence of the kicked rotor to select narrow
parts of an initial momentum distribution. In
refs.\cite{Saunders,Halkyard} the dynamics of a dilute atomic gas
kicked periodically has been studied. There, the authors obtain
analytical expressions for the time evolution of the moments of
momentum for extended initial conditions, corresponding to the
Boltzmann distribution of an ideal gas.

On the other hand, in a recent paper \cite{German}, the evolution of
initially extended distributions in the quantum walk (QW) on the
line was studied. In that work, through an analysis of the
dispersion relation of the process and the introduction of extended
initial conditions, continuous wave equations are derived. In
particular, for a class of initial conditions, the evolution is
dictated by the Schr\"{o}dinger equation of a free particle. This
allows to devise an initially extended condition leading to a
uniform probability distribution whose width increases linearly with
time, and with
increasing homogeneity. Additionally, in some previous works \cite%
{alejo1,alejo2,alejo5,alejo6,auyuanet,victor,monva} a parallelism
between the behavior of the QKR and a generalized form of the QW was
developed, showing that these models have similar dynamics. In the
same line of thought of reference \cite{German}, in this paper we
study the dependence of the QKR in resonance with extended initial
conditions. In particular we obtain, analytically, a dispersion
relation that controls not only the velocity of the wavepacket, but
also the evolution of its shape as a function of time.

The paper is organized as follows: In the next section we present a
brief revision of the QKR equations, in the third section we present
the dispersion relation of the map and the group velocity, in
section $4$ we study the behavior of the system under extended
initial conditions, in section $5$ numerical results are presented
for the secondary resonances, and in the last section we draw the
conclusions.

\section{Quantum kicked rotor}

The QKR is one of the simplest and best investigated quantum models whose
classical counterpart displays chaos. The Hamiltonian has the following
shape
\begin{equation}
\mathbf{H}(t)=\frac{\mathbf{L}^{2}}{2I}+K \cos \mathbf{\Theta}
\sum_{n=1}^{\infty }\delta (t-nT),  \label{qkrham}
\end{equation}
where $K$ is the strength parameter, $I$ is the moment of inertia of the
rotor, $\mathbf{L}$ the angular momentum operator, $T$ the kick period and $%
\mathbf{\Theta}$ the angular position operator. Given the delta function
appearing in eq.(\ref{qkrham}), the external kicks occur at times $t=nT$
with $n$ integer. The index $n$ will be equivalent to time in units of $T$
and the two operators $\mathbf{L}$ and $\mathbf{\Theta}$ satisfy the
canonical commutation rule
\begin{equation}
[\mathbf{L},\mathbf{\Theta}]=-i\hbar.  \label{commuta}
\end{equation}
In the angular momentum representation, $\mathbf{L}|\mathit{l} \rangle =%
\mathit{l} \hbar |\mathit{l} \rangle $, the wavevector is
\begin{equation}
|\Psi (n)\rangle =\sum_{\mathit{l} =-\infty }^{\infty }a_{\mathit{l} }(n)|%
\mathit{l} \rangle  \label{psi}
\end{equation}
and the angular momentum and energy are
\begin{equation}
L(n)=\left\langle \Psi \right\vert \mathbf{L}\left\vert \Psi \right\rangle
=\hbar \sum_{\mathit{l} =-\infty }^{\infty }\mathit{l} \left\vert a_{\mathit{%
l} }(n)\right\vert ^{2},  \label{momento}
\end{equation}
\begin{equation}
E(n)=\left\langle \Psi \right\vert \mathbf{H}\left\vert \Psi \right\rangle
=\varepsilon \sum_{\mathit{l} =-\infty }^{\infty }\mathit{l} ^{2}\left\vert
a_{\mathit{l} }(n)\right\vert ^{2},  \label{ene}
\end{equation}
where $\varepsilon =\hbar ^{2}/2I$. The evolution operator $U$ is built from
the Hamiltonian eq.(\ref{qkrham}). Then the unitary dynamics of the system
is described by
\begin{equation}
|\Psi (n+1)\rangle=\mathbf{U}|\Psi (n)\rangle.  \label{mapa0}
\end{equation}
Substituting eq.(\ref{psi}) in eq.(\ref{mapa0}) and projecting over the
angular momentum state $|l\rangle$, the dynamical equation for the
amplitudes $a_{\mathit{l} }$ is determined
\begin{equation}
a_{\mathit{l} }(n+1)=\sum_{j=-\infty }^{\infty }U_{\mathit{l} j}a_{j}(n),
\label{mapa}
\end{equation}%
where the matrix elements of operator $\mathbf{U}$ are \cite{CCI79}
\begin{equation}
U_{\mathit{l} j}=i^{-(j-\mathit{l} )}e^{-ij^{2}\varepsilon T/\hbar }\,J_{j-%
\mathit{l} }(\kappa),  \label{evolu}
\end{equation}%
$J_{\mathit{l}}$ being the $\mathit{l}$th order cylindrical Bessel function
and its argument the dimensionless kick strength $\kappa\equiv K/\hbar$. The
resonance condition does not depend on $\kappa$ and takes place when the
frequency of the driving force is commensurable with the frequencies of the
free rotor. Inspection of eq.(\ref{evolu}) shows that the resonant values of
the scale parameter $\tau\equiv \varepsilon T/\hbar$ are the set of the
rational multiples of $2\pi$, $\tau=2\pi$ $p/q$. In what follows we assume
that the resonance condition is satisfied and therefore the evolution
operator depends on $\kappa$, $p$ and $q$. We call a resonance primary when $%
p/q$ is an integer and secondary when it is not.

\section{The dispersion relation and the group velocity}

We shall consider the primary resonances $p/q=1$ and search for the solution
of eq.(\ref{mapa}) with the help of discrete Fourier analysis. We define the
discrete Fourier transform of $a_{\mathit{l} }(n)$ as
\begin{equation}
\widetilde{a}_\theta(n)\equiv \sum_{l}e^{i(\theta \mathit{l}+\omega n)}a_{%
\mathit{l} }(n),  \label{transfor}
\end{equation}
where $\omega$ will be determined as a function of $\theta$. Eq.(\ref%
{transfor}) can be inverted to obtain
\begin{equation}
a_{\mathit{l} }(n)=\frac{1}{2\pi }\int_{-\pi }^{+\pi } \widetilde{a}%
_\theta(n) e^{-i(\theta \mathit{l}+\omega n)} d\theta.  \label{antitransfor}
\end{equation}
Substituting eq.(\ref{antitransfor}) in eq.(\ref{mapa}) and using the
identity \cite{Gradshteyn}
\begin{equation}
e^{-i\kappa\cos \theta}=\sum_{j=-\infty}^{\infty}i^{-j}e^{-i\theta
j}\,J_{j}(\kappa),  \label{inter}
\end{equation}
the following dispersion relation between $\omega$ and $\theta$ is obtained
\begin{equation}
\omega=\kappa \cos \theta.  \label{dispersion}
\end{equation}
Therefore, the group velocity of the system is
\begin{equation}
v_g\equiv\frac{d\omega}{d\theta}=-\kappa \sin \theta.  \label{vgrupo}
\end{equation}
Eqs.(\ref{dispersion}) and (\ref{vgrupo}) allow us to make simple
but relevant predictions about the QKR dynamics. In particular, from
eq.(\ref{vgrupo}) it results that the group velocity will be maximum, $%
v_g=\kappa$, if $\theta=\frac{2m+1}{2}\pi$ and minimum, $v_g=0$, if $%
\theta=m\pi$ with $m=0,\pm1,\pm2,...$. If the initial state is a wavepacket
close to some eigenfunction
\begin{equation}
a_{\mathit{l} }(0)=f(\mathit{l}) e^{i\theta_0 \mathit{l}},  \label{plana}
\end{equation}
where $f(\mathit{l})$ is a smooth function of $\mathit{l}$ and the value of $%
\theta_0$, the initial angular position, determines the velocity of
the wavepacket. If $\theta_0=0$ a sufficiently extended wavepacket
should stay at rest because $v_g= 0$, while if $\theta_0=\pm \pi/2$
it should move with maximum velocity $v_g=\pm \kappa$. In order to
verify this statement, in the next section we choose the initial
condition as
\begin{equation}
a_{\mathit{l}}(0)=\left[ \frac{1}{\sigma_0\sqrt{2\pi }} \exp{\ (-\frac{%
\mathit{l}^2}{2{\sigma_0}^2})} \right]^{\frac{1}{2}} e^{i \theta_0 \mathit{l}%
} ,  \label{aes}
\end{equation}
where $\sigma_0$ is the standard deviation of the initial
distribution. We shall then calculate the average momentum and
energy obtained from this initial condition.

\section{Moments with extended initial conditions}

In the case of the primary resonances, eq.(\ref{mapa}) is solved
\cite{CCI79} using the recursion relation satisfied by the Bessel
functions. The general solution can be written as
\begin{equation}
a_{\mathit{l}}(n)=\sum\limits_{j=-\infty }^{\infty }\left( -i\right) ^{%
\mathit{l}-j}a_{j}(0)\,J_{\mathit{l}-j}(n\kappa),  \label{solua}
\end{equation}%
where $a_{j}(0)$ are the initial amplitudes. The probability distribution at
time $n$ is given by
\begin{equation}
P_{\mathit{l}}(n)=|a_{\mathit{l}}(n)|^{2},  \label{prob}
\end{equation}%
that can be expressed as
\begin{equation}
P_{\mathit{l}}(n)=\sum_{j,k=-\infty }^{\infty }\left(-i\right) ^{k-j}
a_{k}(0)a_{j}^{\ast }(0) J_{\mathit{l}-k}(\kappa n)J_{\mathit{l}-j}(\kappa
n).  \label{prob1}
\end{equation}%
All the moments of $P_{\mathit{l%
}}(n)$ can be calculated analytically from eq.(\ref{prob1}) using
the properties of the Bessel functions. We calculate the first and
second statistical moments
\begin{equation}
M_{1}(n)=\sum_{j=-\infty }^{\infty }j\left|a_{j}(n)\right|^2 ,  \label{mo0}
\end{equation}
\begin{equation}
M_{2}(n)=\sum_{j=-\infty }^{\infty }j^2\left|a_{j}(n)\right|^2,  \label{mo2}
\end{equation}
obtaining:
\begin{equation}
M_{1}(n)=-\kappa n\sum_{j=-\infty }^{\infty }\Im \left[ a_{j}(0)a_{j-1}^{%
\ast }(0)\right] +M_{1}(0),  \label{mome0}
\end{equation}
\begin{eqnarray}
M_{2}(n)&=&\frac{(\kappa n)^{2}}{2}\left( 1-\sum_{j=-\infty }^{\infty }\Re %
\left[ a_{j}(0)a_{j+2}^{\ast }(0)\right] \right)  \nonumber \\
&&+\kappa n\sum_{j=-\infty }^{\infty }\left( 2j+1\right) \Im \left[
a_{j}(0)a_{j+1}^{\ast }(0)\right]  \nonumber \\
&&+M_{2}(0),  \label{mome}
\end{eqnarray}%
where $\Re \left[ x\right]$ and $\Im \left[ x\right]$ are
respectively the real and imaginary part of $x$. We note that the
first and the second moments are related to the average angular
momentum and the average energy of the system respectively
\begin{equation}
M_{1}(n)=\frac{L(n)}{\hbar},  \label{mome002}
\end{equation}
\begin{eqnarray}
M_{2}(n)=\frac{E(n)}{\varepsilon} .  \label{mome0002}
\end{eqnarray}
Substituting eq.(\ref{aes}) into eqs.(\ref{mome0}, \ref{mome}) and using
that
\begin{equation}
\sum_{j=-\infty }^{\infty }\exp \left( -\frac{j^{2}}{2\sigma^{2}}%
\right)\cong \sqrt{2\pi} \sigma ,  \label{ayuda}
\end{equation}
for $\sigma>1$, the following analytical results for the moments are
obtained
\begin{equation}
M_{1}(n)=-\kappa n \sin \theta_0\ e^{-\frac{1}{8\sigma_0}},  \label{mome02}
\end{equation}
\begin{eqnarray}
M_{2}(n)=\frac{(\kappa n)^{2}}{2}\left[ 1-\cos {2\theta_0}\ e^{-\frac{1}{%
2\sigma_0}} \right]+M_{2}(0),  \label{mome2}
\end{eqnarray}
and the variance, $\sigma^2 \equiv{M_{2}-M_{1}^{2}}$, is then
\begin{eqnarray}
\sigma^2(n)&=&\frac{(\kappa n)^{2}}{2}\left[ 1-\cos {2\theta_0}\ e^{- {\frac{%
1}{2\sigma_0}}}-2\ {\sin^2 \theta_0}\ e^{{-\frac{1}{4\sigma_0}}} \right]
\nonumber \\
&&+\sigma_{0}^2.  \label{variance}
\end{eqnarray}
Eq.(\ref{mome2}) shows a sinusoidal dependence of the energy with $\theta_0$%
, the angular initial position for the initial distribution. This
dependence seems to be confirmed in the experimental realization of
ref.\cite{Currivan}. From the time dependence of
eq.(\ref{variance}), it is clear that the ballistic behavior of the
resonance is independent of
the initial condition. However, from this equation we also conclude that, for $n$ fixed, $%
\sigma^2(n)\mapsto {\sigma_0}^2$ when $\sigma_0\gg1$. This means
that when $\sigma_0$ is very large the system can take a large time
to exhibit its ballistic behavior. Therefore, the experimental
observation of this behavior will be subordinated to the extension
of the initial distribution. This result could explain the absence
of ballistic growth in some experimental realizations
\cite{Bharucha}. Additionally, our previous reasoning can also be
related to refs.\cite{Saunders,Halkyard} where the authors find
analytical expressions for the moments with extended initial
conditions that correspond to the Maxwell-Boltzmann ideal gas
distribution at finite temperatures. The parameter $\sigma_0$ in our
formulation can be qualitatively related to the temperature, a high
temperature corresponding to a high value of $\sigma_0$ and in both
formulations the behavior of the moments is consistent.

\begin{figure}[h]
\resizebox{0.4\textwidth}{!}{%
  \includegraphics{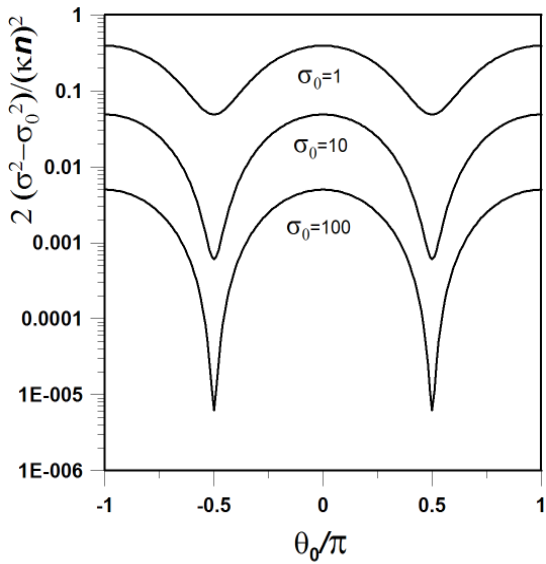}
}
\caption{Using eq.(\protect\ref{variance}) the function $2\frac{\protect%
\sigma^2-{\protect\sigma_0}^2}{(\protect\kappa n)^2}$ is presented, in a
log-scale, as a function of the initial angular position, $\protect\theta_0/%
\protect\pi$, for three values of $\protect\sigma_0$: $1$, $10$ and $100$.
The calculation corresponds to the primary resonance $p/q=1$.}
\label{f1}
\end{figure}

\begin{figure}[h]
\resizebox{0.4\textwidth}{!}{%
  \includegraphics{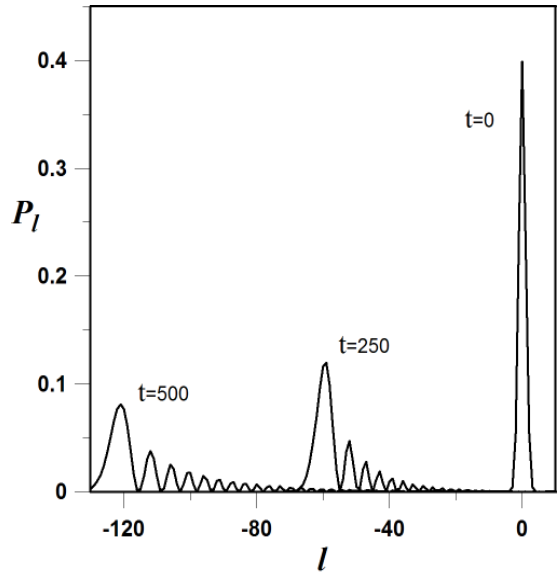}
} \caption{The probability distribution in momentum space as a
function of the dimensionless angular momentum is presented for
three times $t=0$, $t=250$ and $t=500$. The figure was obtained
numerically from the map eq.(\protect \ref{mapa}) with
$\protect\kappa=0.25$ in the primary resonance $p/q=1$. The
initial conditions have been given by eq.(\protect\ref{aes}) with $\protect%
\sigma_0 = 1$ and $\protect\theta_0=\protect\pi/2$.}
\label{guzman}
\end{figure}

From eq.(\ref{mome02}) and eq.(\ref{mome002}) the QKR group velocity is
calculated
\begin{equation}
v_g=\frac{dM_{1}(n)}{dn}=\frac{1}{\hbar}\frac{dL(n)}{dn}=-\kappa \sin
\theta_0\ e^{{-\frac{1}{8\sigma_0}}},  \label{vgrupo2}
\end{equation}
and we confirm that if $\theta_0=0$ then $v_g= 0$, while if
$\theta_0=\pm \pi/2$ then $v_g\simeq\pm \kappa$ for $\sigma_0>1$.
This result is very interesting because it shows that if one chooses
the value of $\theta_0$ appropriately in the initial condition of
the QKR, one can get an average angular momentum pointing in a
preset direction and increasing in time with a preset ratio $v_g$.
In order to visualize the temporal evolution of the probability
distribution in momentum space, we present the fig.\ref{guzman}
obtained numerically using the original map eq.(\ref{mapa}). There,
the initial conditions $\theta_0=\pi/2$ in eq.(\ref{aes}) have been
chosen so that the group velocity is negative, hence, the first
moment moves to the left. The spreading of the wave function in the
angular momentum space as time elapses can also be appreciated in
the figure.

Strictly, from eq.(\ref{variance}) and fig. \ref{f1} the values of
$\theta_0$ and $\sigma_0$ cannot be chosen in order to produce
$\sigma(n)=\sigma_0$ with $v_g\neq0$. However, for $\theta_0=\pm
\pi/2$ and $\sigma_0>1$ it is possible to produce a wavepacket with
a constant shape that travels with constant velocity for a very long
time. Additionally, in the asymptotic limit
$\sigma_0\rightarrow\infty$, from eq.(\ref{vgrupo2}) we recover the
expected equation eq.(\ref{vgrupo}) that corresponds to the group
velocity for a train of plane waves.

Therefore, we have demonstrated that the dispersion relation eq.(\ref{vgrupo}%
) controls the velocity of the QKR wavepacket and the temporal evolution of
its shape, see eqs.(\ref{mome02}, \ref{mome2}).
\begin{figure}[h]
\resizebox{0.4\textwidth}{!}{%
  \includegraphics{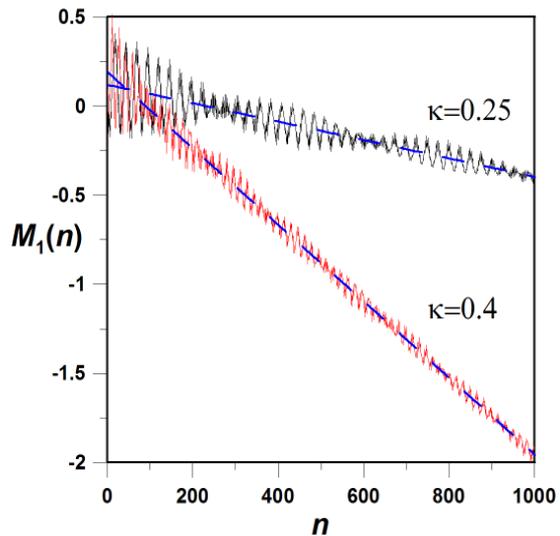}
}\caption{The first statistical moment as a function of the
dimensionless time $n$ for two values of $\protect\kappa=0.25$ and
$0.4$. The dashed lines
are lineal adjustment, the angular initial position is $\protect\theta_0=%
\protect\pi/2$ and $\protect\sigma_0=1$. The calculation corresponds to the
secondary resonance $p/q=1/3$.}
\label{f2}
\end{figure}

\section{Secondary resonances}

We now study the incidence of extended initial distributions on the
secondary resonances. In this case $p/q$ is not an integer and eq.(\ref%
{evolu}) becomes too complicated to handle analytically, an
exception being the antiresonance case $p/q=1/2$ \cite{Izrailev}.
Eq.(\ref{mapa}) is no longer invariant under translations in the
angular momentum. This invariance was fundamental for the previous
analytical treatment. We therefore study  numerically the influence
of the extended initial conditions in the dynamics of the secondary
resonances, using eq.(\ref{mapa}) with $p/q=1/3$, and initial
conditions given by eq.(\ref{aes}).

\begin{figure}[h]
\resizebox{0.4\textwidth}{!}{%
  \includegraphics{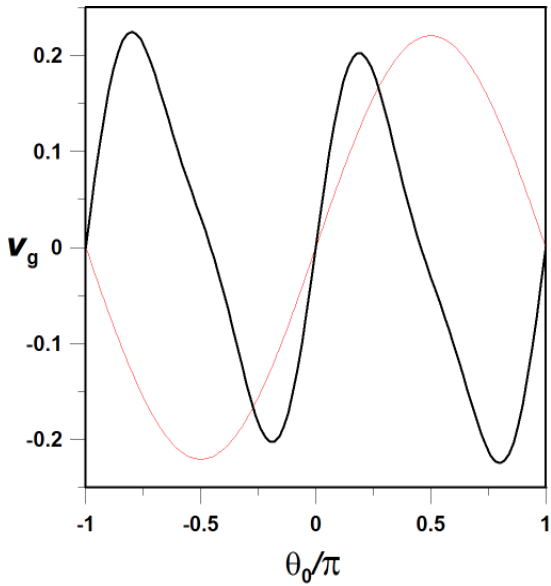}
} \caption{The dimensionless $v_g$ as a function of the initial
angular position $\protect\theta_0$ is presented. A thin red line is
used for the primary resonance and a thick black line for the
secondary resonance $p/q=1/3 $. Both calculations have
$\protect\sigma_0=1$ and $\protect\kappa=0.25$ but $v_g$ for the
secondary resonance is multiplied by a factor of $60$.} \label{f3}
\end{figure}

\begin{figure}[h]
\resizebox{0.4\textwidth}{!}{%
  \includegraphics{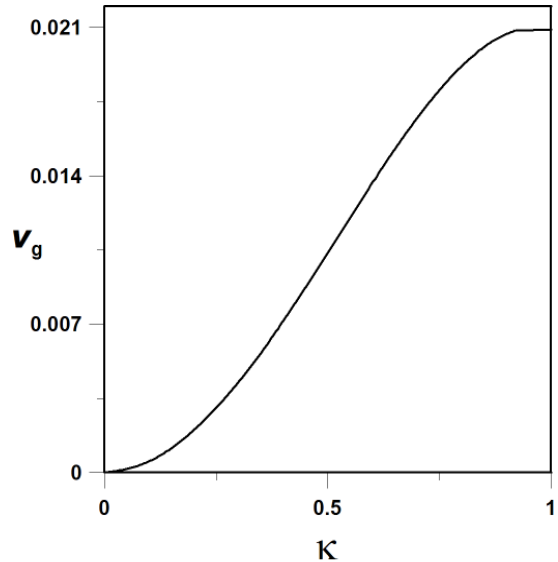}
}\caption{The dimensionless $v_g$ as a function of the strength parameter $%
\protect\kappa$ for the secondary resonance $p/q=1/3$ with $\protect\sigma%
_0=1$ and $\protect\theta=\protect\pi/4.$}
\label{f4}
\end{figure}

In fig. \ref{f2} we present the numerical calculation of the first
statistical moment $M_{1}(n)$ as a function of time $n$, for two
different values of $\kappa$ with $\theta_0=\pi/2$ and $\sigma_0=1$.
This figure shows a complex oscillatory behavior where it is
possible to define an average lineal behavior, shown by the blue
dashed lines. The slopes of these straight lines define the
corresponding group velocities for the secondary resonances. In fig.
\ref{f3} we present the numerical calculation of the group velocity
of the system as a function of the angular initial position. This
figure shows a quasi sinusoidal dependence of $v_g$ with $\theta_0$,
that reminds us of the behavior that was found for the primary
resonances, eq.(\ref{vgrupo2}). However, if this behavior is
compared to that of the previous case, now the oscillation frequency
duplicates its value and its
amplitude decreases by some orders of magnitude (note that this curve in fig. \ref%
{f3} is multiplied by $60$). Additionally, fig. \ref{f2} also shows that $v_g
$ has a dependence with the strength parameter $\kappa$. We have obtained
numerically this dependence and we synthesize the result in fig. \ref{f4}.
From this figure, it is clear that the simple proportionality obtained for
the primary resonance, see eq.(\ref{mome02}), has been lost.


\section{Conclusion}

In summary, through the use of the discrete Fourier transform in the
study of the primary resonances of the QKR, it is straightforward to
get an explicit dispersion relation. This dispersion relation is a
powerful tool for predicting the QKR dynamics because it controls
not only the velocity of the wavepacket, but also the evolution of
its shape as time elapses. We show that the extension of the initial
distribution can delay the characteristic time that the resonant
evolution needs in order to exhibit its ballistic behavior. This
result could explain the absence of the ballistic growth in some
experimental realizations of the QKR. Additionally, from the
analytical treatment we show the sinusoidal dependence between the
energy and the initial average position of the system.

We also study, numerically, the influence of an extended initial
distribution in the behavior of the secondary resonances of the QKR.
We obtain a relation between the parameters of the extended initial
condition and the dynamical average of the system. We show, as in
the case of the primary resonances, that if one chooses the value of
$\theta_0$ appropriately in the initial condition, one can get an
average angular momentum pointing in a preset direction and
increasing in time with a preset ratio $v_g$.

\thanks{We acknowledge the support from PEDECIBA, CSIC,
ANII and thank V. Micenmacher, E. Rold\'{a}n and G. J. Valc\'{a}rcel
for their comments and stimulating discussions.}

\end{document}